\documentclass[12pt]{article}
\usepackage{epsfig}
\usepackage{hhline}
\def\x {\times}

\def\be {\begin{equation}}
\def\ee {\end{equation}}
\def\bea {\begin{eqnarray}}
\def\eea {\end{eqnarray}}
\def\pd {\partial}
\def\tfrac#1#2{{\textstyle{#1\over #2}}}
\def\x {\times}
\def\nn {\nonumber}
\def\tfrac#1#2{{\textstyle{#1\over #2}}}

\begin{document}
\renewcommand{\arraystretch}{1.5}

\rightline{UG-9/96}
\rightline{hep-th/9612095}
\rightline{December 1996}
\vspace{1.5truecm}
\centerline{\bf  MULTIPLE INTERSECTIONS OF $D$-BRANES AND $M$-BRANES}
\vspace{1.5truecm}
\centerline{\bf E.~Bergshoeff, \ M.~de  Roo, \  E.~Eyras,}
\centerline{\bf B.~Janssen \ and \ J.~P.~van der Schaar}
\vspace{.4truecm}
\centerline{{\it Institute for Theoretical Physics}}
\centerline{{\it Nijenborgh 4, 9747 AG Groningen}}
\centerline{{\it The Netherlands}}
\vspace{3truecm}
\centerline{ABSTRACT}
\vspace{.5truecm}
We give a classification of all multiple intersections 
of $D$-branes in ten dimensions and $M$-branes in eleven dimensions
that correspond to threshold BPS bound states.
The residual supersymmetry of these composite branes is determined.
By dimensional reduction  composite $p$-branes in lower dimensions can
be constructed. 
We emphasize in dimensions $D\ge 2$ those solutions which involve
 a single 
scalar and depend on a single harmonic function. For these
extremal branes we obtain the strength of the coupling between
the scalar and the gauge field.
In particular, we give a $D$-brane and $M$-brane 
interpretation of extreme $p$-branes in two, three and four dimensions.
\newpage

\section{Introduction}

\noindent
Classical solutions of the low-energy effective supergravity action  have
 played a crucial role in the unification of string
 theories. This unification, in terms of the conjectured $M$-theory
 \cite{Town,Witt},
 has led to a renewed interest in $D=11$ supergravity. 
 Since $D=11$ supergravity is the low-energy limit of $M$-theory,
 its classical solutions, and their descendants which can be obtained in
 lower dimensions by dimensional reduction, play a particularly important
 role. The basic eleven-dimensional extended objects ($Mp$-branes
or, shortly, $M$-branes) are the
 $M2$-brane \cite{Duff-Stelle} and the $M5$-brane \cite{Guven}.
 
In ten dimensions a
 class of solutions of Type IIA/IIB string theory, 
 satisfying Dirichlet boundary conditions in certain directions,
 has received much attention \cite{Pol1,Pol2}.
 These solutions are called Dirichlet $p$-branes or shortly $Dp$-branes (or
just $D$-branes). 
 The charge of the $Dp$-branes is carried by a Ramond-Ramond
 gauge field. $Dp$-branes exist for all values of $0\le p\le 9$ and
 are all related by $T$-duality \cite{Pol2,Pol3,Bachas,Alvarez,Be1}.

Both the $M$-brane and the $D$-brane solutions are characterized by a function
 $H$ which depends only on the coordinates transverse to the brane, 
 and is harmonic 
 on this transverse space. 
 This suggests that the $M$-brane and $D$-brane solutions are related, 
and indeed
 one finds that direct dimensional reduction of the $M2$-brane 
 and double dimensional reduction of the $M5$-brane in $D=11$ leads
 to the $D2$- and $D4$-branes in IIA supergravity. 

It is natural to consider solutions that correspond to 
intersections of $D$-branes and $M$-branes.
This was first done in \cite{Papad-Town} where certain solutions occurring in 
\cite{Guven} were interpreted as intersections of $M$-branes. Soon after,
these results were generalized both for $M$-branes 
\cite{Tseyt1,Klebanov,Gaunt,Khviengia,Costa,Pap} as well as $D$-branes 
\cite{Tseyt1,BeBeJa,Gaunt,Papad-Town2}. In the latter case the solutions
correspond to bound states of $D$-branes \cite{Pol2,Green}.

Under certain conditions the force between two $M$-branes
 or $D$-branes vanishes \cite{Tseyt2}, and 
 composite configurations 
 of branes can be static solutions to the equations of motion. 
 The conditions for the existence of such configurations have been formulated. 
In particular, the so-called 
 harmonic function rule \cite{Tseyt1}
 prescribes how products of powers of the
 harmonic functions $H_i$ of the $N$ intersecting branes must occur in
 the composite solution. In particular, it implies that if one of the
 $H_i$ is set equal to one, a solution with $N-1$ intersecting branes
 is obtained.

For a more detailed analysis of configurations of two branes,
 we assume that the powers of harmonic functions are as stated by the
 harmonic function rule, and then split the coordinates in three
 parts: the overall world-volume coordinates ($d$), which are common to the
 two branes, the overall transverse coordinates ($t$), 
 and the remainder, which
 are called relative transverse ($n$), and are 
 transverse to only one of the two
 branes. Three kinds of intersections of a $p_1$- and a $p_2$-brane
 are possible, with the following
 conditions,
 valid for $D$-branes in $D=10$ and
 the basic $M$-branes in $D=11$,
 on $H_1$ and $H_2$ \cite{Pol2,Tseyt1,Gaunt,Tseyt2,BeBeJa}:

\begin{enumerate}
\item Both $H_i$ depend only on the $t$ overall transverse coordinates. Then
 in $D=10$ we must have $n=4$ (i.e.~4 relative transverse directions), 
while in $D=11$ 
 the only possibilities are 
 $(0|2,2)$ and $(3|5,5)$ $(n=4)$, and $(1|2,5)$ 
$(n=5)$\footnote{We denote the intersection of a $p_1$- and a $p_2$-brane
over a common $q$-brane with $d = q+1$ by $(q|p_1,p_2)$.
 We do not include the case $n=0$ $(p_1=p_2)$, 
 for which the
 intersection is described by the sum of $H_1$ and $H_2$ (multiple
 branes of the same type with different locations).}.

\item One $H_i$ depends on the overall transverse coordinates, the other
 on the relative coordinates. In this case the conditions on $n$ are as
 in case (1).

\item Both $H_i$ depend on the relative coordinates. Then in $D=10$ and
 $D=11$ we must have $n=8$, which in $D=11$ can only be realized 
 as $(1|5,5)$ \cite{Gaunt}.

\end{enumerate}

\noindent
In case more than two branes intersect, the above rules must apply for
 each pair of branes in the composite system. As we will see, this basic
requirement enormously restricts the number of allowed multiple
intersections.

The aim of this paper 
 is to give a systematic 
 and complete classification of intersecting branes in ten and eleven 
 dimensions {\it satisfying the above conditions}. This paper will
mainly concentrate on the intersecting branes that satisfy condition (1).
The ones that satisfy conditions (2) or (3) are separately discussed in
an Appendix, since their status in string theory is less clear.

Note that the harmonic
 function rule implies restrictions on the form of the metric of the
 solution, and that therefore our analysis does not exclude the existence 
 of other (static) multiple brane configurations. In particular, we
only consider intersections where each participating brane corresponds to
an independent harmonic function in the solution. Such solutions
correspond to threshold BPS bound states, i.e. they satisfy the
no force condition. To summarize, in this paper we only consider
solutions corresponding to threshold BPS bound states. We do {\it not}
consider the following solutions (see, however, the Conclusions):

\begin{itemize}

\item $D=10$ intersections that involve NS-NS
       strings, five-branes and/or their $T$-duals. Neither do we
       consider in $D$=11 intersections that involve a gravitational wave 
       (boosted $M$-branes \cite{Tseyt3}) and/or
       its magnetic partner. 

\item Solutions corresponding to non-threshold BPS bound states such as
the $D=10$ $D$-brane bound states with $n=2$ or $6$ \cite{Pol2},
the $D=10$ $(q_1,q_2)$ string solutions of \cite {Schwarz} and the
$D=10$ ($D=11$) solution given in \cite{BBO} (\cite{To})
       that interpolates between a 2-brane and a 5-brane. 

\end{itemize}

Our main conclusions are that under condition (1) there are three 
 inequivalent ways of eight intersecting $p$-branes, both in
 $D=10$ and $D=11$. If intersections with $n=8$ are allowed as well,
 a ninth brane can be added to these configurations.

A single $M$- or $D$-brane preserves half the supersymmetry of the
 corresponding supergravity theory (which has 32 real supersymmetry 
 generators). As a general rule, 
 a configuration of $N$ intersecting branes
 preserves {\it at least} $1/2^N$ of the maximal supersymmetry 
 \cite{Pol2, Tseyt1}. In our analysis we will see when and how the
 ``at least'' becomes relevant: in some cases an additional brane
 can be added to a composite system without additional breaking of
 supersymmetry (\cite{Green,Klebanov,Gaunt}). Our maximal intersecting
 configurations with eight branes preserve $1/32$ or $1/16$ of the maximal
 supersymmetry, for the intersection with nine branes this is $1/32$.

The conditions (1-3) follow from the 
 gauge field equation of motion, and so are {\it a priori} necessary
 conditions. The Einstein equation, and, in $D=10$, the dilaton
 equation of motion, need to be checked. In all cases considered
 in this paper we find that the full equations of motion are
 satisfied by multiple intersections satisfying the above
 conditions. It is an interesting fact that all multiple
 configurations based on (1-3) preserve at least $1/32$ of the
 maximal supersymmetry. The reason must be that the condition of
 vanishing force between branes is implied by supersymmetry. If
 this is true, then on the one hand it should be possible to derive
 the conditions (1-3) from the requirement of supersymmetry, while
 on the other hand preservation of supersymmetry should imply the
 complete equations of motion. However, we have not proved this.
 
The organisation of this paper is as follows. In the body of the paper
 we will extensively discuss case (1) of the above conditions:
 $n=4,5$ and dependence on the overall transverse coordinates. We will
 do this for $D=10$ in Section 2, and for $D=11$ in Section 3.
 In Section 4 we discuss some aspects of the reduction of
 our composite solutions to lower dimensions, with emphasis on extreme
$p$-branes in 2,3 and 4 dimensions.
 Further remarks, in particular on the inclusion of NS-NS branes and/or
non-threshold BPS bound states, are
 made in the Conclusions.
 Explicit
 representations of the solutions with maximum numbers of
 intersecting branes are given in Appendix A and B for $D=10$ and $D=11$, 
 respectively.
 In  Appendix C we discuss the additional possibilities which
 arise from the cases (2), (3) of the above conditions. 
\section{Intersections of $D$-branes in 10 dimensions}

\noindent
The single elementary Dirichlet $p$-brane solution in the string
 frame in ten dimensions is given by the following  metric, dilaton
 and gauge field\footnote{We use here the basis of RR gauge fields that
occur naturally in the Wess-Zumino terms of the
$D$-brane actions \cite{GHT}. In this basis the
Chern-Simons terms in the curvatures for the RR gauge fields always
contain a NS-NS 3-form curvature. These CS terms vanish
for the class of solutions we are considering in this paper.}:
\begin{eqnarray}
ds^2 = H_p^{-1/2}\  ds^2_{p+1} - 
{H_p^{1/2}} \ ds^2_{9-p}, \nonumber \\
e^{2\phi}= (H_p)^{-\frac{1}{2}(p-3)}, \ \  F_{01...pi}= \partial_i H_p^{-1}, 
\label{singlebrane}
\end{eqnarray}
where $H_p$ is a harmonic function which depends on the $9-p$
 transverse coordinates\footnote{Sofar we used a notation where 
the subscript $i$ on the 
harmonic function indicated the $i$th intersecting $D$-brane.
Sometimes, however, like here, it is more convenient to use a notation 
where the subscript $i$ on the harmonic function indicates the number 
$p$ of the corresponding $p$-brane. It should be clear
from the context which of the notations is used.}. The line element 
$ds^2_{p+1}$ contains the time coordinate $t=x^0$.

$T$-duality transforms a $p$-brane into a $(p+1)$-brane if $H_p$ is
 independent of one of the transverse coordinates, say $x$. The
 duality rule for the metric is given
 by \begin{equation} {\tilde g}_{xx}=1/g_{xx}, \end{equation} so
 that a transverse direction gets dualized to a worldvolume
 direction. The other rules of $T$-duality can be found
 in \cite{BeHuOr,Be1}. It is in principle possible to
 perform $T$-duality in the opposite direction and change a worldvolume
 coordinate into a transverse one. However, this is
 a ``dangerous'' $T$-duality, since we have to suppose that the
 harmonic function after dualization depends on the direction 
in which we have dualized and it is not guaranteed that the result is still a
 solution of the equations of motion. In this paper we will perform only
``safe'' $T$-duality transformations. 

It is convenient to 
represent every coordinate that corresponds to a worldvolume
 direction by $\x$ and every direction transverse to the brane
 by $-$. We thus obtain the following representation of the metric of
 a $D$-brane solution: 
\begin{equation}
\label{notation}
ds^2=\underbrace{\x \ \x \ ... \ \x}_{p+1}\  \overbrace{- \ - \ ... \ -}^{9-p}.
\end{equation}
It is easy to see that acting with a ``safe'' $T$-duality on
 this metric, a $-$ changes into a $\x$. 

The Ansatz we use to describe intersecting $D$-branes
 follows from the harmonic function rule: the metric
 is diagonal and every $dx^2$ gets multiplied by a factor which is
 a product of the powers of the harmonic functions involved in the
 intersection. $D$-branes that have the coordinate 
 $x$ as a worldvolume direction
 contribute a factor\footnote{These factors are given for the
 string frame.}
 $H^{-1/2}$ and the ones that have $x$ as 
 a transverse direction contribute $H^{1/2}$. The dilaton is
 given by the product of the dilaton expressions for each
 separate brane and we have a gauge field of the
 form given in (\ref{singlebrane}) for every $D$-brane in the intersection.

As an example we give the expression for the metric and dilaton of
 a $(p+r)$-brane intersecting with a $(p+s)$-brane over
 a $p$-brane, i.e. a $(p|p+r,p+s)$ configuration:  
\begin{eqnarray}
\label{general}
ds^2 &=&\left(H_{p+r}H_{p+s}\right)^{-1/2}ds^2_{p+1} - \left({H_{p+r}\over 
H_{p+s}}\right)^{1/2} ds^2_s   \nonumber \\
&&- \left({H_{p+s}\over H_{p+r}}\right)^{1/2}
ds^2_r -  \left(H_{p+r}H_{p+s}\right)^{1/2} ds^2_{9-p-r-s} \ ,\\
e^{2\phi} &=& e^{-\frac{1}{2}(p+r-3)}e^{-\frac{1}{2}(p+s-3)}\, .\nonumber
\end{eqnarray}
In this case $d=p+1$, $n=r+s$ and $t=9-p-r-s$. In this Section we will
only consider intersections that  satisfy condition (1) 
of the Introduction, the other two cases will be discussed in Appendix C.
For such intersections, the
 two harmonic functions depend on the $t$ overall transverse
 coordinates, and we must have $n=4$ in order that the intersecting
configuration is a solution to the equations of motion. The expressions
for the gauge fields for this case are given by

\begin{equation}
F_{0\cdots p1\cdots ri} = \partial_i H_{p+r}^{-1}\, ,\hskip 1.5truecm
F_{0\cdots p1\cdots si} = \partial_i H_{p+s}^{-1}\, .
\end{equation}

Using the notation of (\ref{notation}) we can rewrite the
general $N=2$ intersection given in (\ref{general}).
For example, a possible two-intersection is given by
\begin{equation}
ds^2= \left\{
\begin{array}{cccccccccccc}
\x&\x&\x&\x&\x&-&-&-&-&-& &:H_1  \\
\x&-&-&\hbox to 0pt{\hss$\underbrace{\hskip2.2cm}_{x_n}$\hskip 3mm\hss}-&
 -&-&-&\hbox to 0pt{\hss$\underbrace{\hskip1.2cm -\hskip 1.2cm}_{x_t}$\hss}
 &-&-& &:H_2.
\end{array} \right.
\label{N=2}
\end{equation}
This is a $(0|0,4)$-solution, i.e.~a 0-brane lying in a
 4-brane, with $d=1,\ n=4,\ t=5$. The harmonic functions $H_i$ both
 depend on the coordinates $x_t$.

In an intersection such as (\ref{N=2}) $T$-duality acts on a
 column, changing every $\x$ in a $-$ and vice versa. If we act
 with $T$-duality on the relative transverse directions and on the
 overall transverse direction we recover the other $T$-dual
 solutions with $n=4$ relative transverse directions given
 in \cite{BeBeJa,Gaunt}. Clearly $n$, the number of relative
 transverse directions, is a $T$-invariant quantity. It is also
 clear that (\ref{N=2}) represents the complete $N=2$ $n=4$ family,
 since all other members can be obtained from it by
 ``safe'' $T$-duality transformations. For the same reason we can limit
 ourselves, for arbitrary $N$, to those intersections with $d=1$
 (intersections over a 0-brane).

When adding further $D$-branes to (\ref{N=2}), in the form of
 horizontal lines with $\x$'s and $-$'s, we need to have $n=4$
 (four different entries of $\x$ or $-$)
 for every pair in the intersection. 
 To streamline the
 construction, it is useful to characterize an intersection by the
 contents of the columns (components of the metric) corresponding
 to the relative transverse coordinates. For an $N$-intersection
 we can, by using $T$-duality, bring each column in a form such
 that no more than $[N/2]$ $\x$'s are present. Such columns
 are the ``building blocks'' of the intersection. Given $N$, there
 are $\left(N\atop k\right)$ building blocks with $k$ $\x$'s. 
 
In an $N$-intersection ($N \ge 2$) there are $\left(N\atop 2\right)$
 intersecting pairs. The total number of differences between $\x$
 and $-$ in the $N$-intersection is therefore equal to 
 $4\left(N\atop 2\right)$. A column with $k$ $\x$'s contributes 
 $k(N-k)$ differences. 
Let $n_k$ be the number of building blocks with $k$ $\x$'s.
 Then we must have
\begin{equation}
\sum_{k=1}^{[N/2]} k (N - k) n_k = 4 \ {N \choose 2}      \, ,
\label{I-form}
\end{equation}
with $\sum_k n_k < 9$.
Given $N$, this is an equation for the $n_k$. 

Let us give a few examples. For $N=2$ there is only one type of
building block with $k=1$. Equation (\ref{I-form}) for this case
reduces to the equation $n_1 =4$ which is condition (1) of the Introduction.
For $N=3$ there is again only one type of building block with $k=1$ and we find
$n_1 = 6$. For $N=4$, there are two types of building blocks, with
$k=1$ and with $k=2$. Equation (\ref{I-form}) reduces to
$3n_1 + 4n_2 = 24$ which has 3 solutions namely
 $(n_1,n_2)=(8,0),(4,3)$ and $(0,6)$. Finally, for $N=5$ there are
again two types of building blocks with $k=1,2$ and we find
$4n_1 + 6n_2 = 40$ leading to 2 solutions given by
$(n_1,n_2)=(4,4)$ and $(1,6)$.
From now on, we will use the numbers $n_k$ to 
 label solutions. Note that the remaining $T$-duality 
and the interchange of columns and/or rows in the
 representation of the metric (corresponding to
a relabeling of the spacetime coordinates or the intersecting branes), 
do not change the $n_k$.

Clearly, (\ref{I-form}) is only a necessary condition for the
 existence of a solution. Given a set of $n_k$ allowed
 by (\ref{I-form}), it is not clear that one can actually realize such a
 solution in terms of the available building blocks and consistent
with condition (1) of the Introduction. In practice,
 we have found that such a realization is possible only in a small
 number of cases. In the actual construction it is not always
 useful to use only building blocks with $k\le [N/2]$. Instead, it is
 convenient to start the $N$-intersection with a 0-brane.
 Since $n=4$, all other branes in the intersection must then be
 4-branes. 
We find that for $N=2,..,5$ all configurations that satisfy the consistency
condition (\ref{I-form}) actually satisfy the stronger condition (1) of
the Introduction for each intersection. However, for $N=6$,
the $(n_1,n_2,n_3) = (3,0,5)$  configuration does not survive. We
have repeated this analysis until we reach $N=8$ with three different
configurations. At this point, our procedure stops. 
Although (\ref{I-form}) has solutions for $N=9$, it 
 turns out to be impossible to add a ninth brane in such a
 way that it has $n=4$ relative transverse directions with all
 other 8 $D$-branes. An overview of
 the different intersections and their relations
 is given in Figure (\ref{D-tree}), the explicit form of
 the three 8-intersections is given in Appendix \ref{N=8}.

A crucial role in the classification given in Figure (\ref{D-tree}) 
is played by the observation that 
up to $T$-duality and interchanges of 
rows and/or columns there is a {\it unique} $D$-brane configuration
that realizes the numbers $(n_1, n_2, ...)$ 
obtained from (\ref{I-form}) and given in Figure ({\ref{D-tree}). 
Sofar, we have not been able to give a general proof of this fact.
Instead, we checked it by brute force in a case by case analysis.

As we mentioned in the Introduction, at this stage one should still
 check the Einstein equation and the dilaton equation of motion.
 We have checked these for the three $N=8$ configurations using
the computer.
 This implies that the intersections with $N\ge 5$ are also
 solutions. For lower $N$ the number $t$ of overall transverse
 coordinates increases, so that the harmonic functions can depend
 on more  coordinates. We checked that the equations of motion
 indeed allow this.

Let us now consider supersymmetry. A single $D$-brane has half of
 its supersymmetry unbroken, since the supersymmetry
 transformations for the Killing spinor in the string frame
\begin{eqnarray}
\delta \psi_\mu &=&
 \partial_\mu\epsilon -\tfrac{1}{4} \omega_\mu^{ab}\gamma_{ab}\epsilon +
 \frac{(-)^p}{8 (p+2)!}\ e^\phi\ F_{\mu_1 ...\mu_{p+2}}
        \gamma^{\mu_1 ...\mu_{p+2}} \gamma_\mu \epsilon_{(p)}^\prime = 0\, ,
\nonumber\\
\delta\lambda &=& 
  \gamma^\mu (\partial_\mu\phi)\epsilon + {3-p\over
  4(p+2)!}\ e^\phi \ 
  F_{\mu_1\cdots \mu_{p+2}} \gamma^{\mu_1\cdots \mu_{p+2}}\ 
  \epsilon_{(p)}^\prime = 0\, ,
\end{eqnarray}
for the $D$-brane yield

\begin{equation}
 \epsilon + \gamma_{01\cdots p}\ \epsilon_{(p)}^\prime = 0\, . 
\label{proj}
\end{equation}

\newpage
\begin{figure}[ht]
\centering
\input{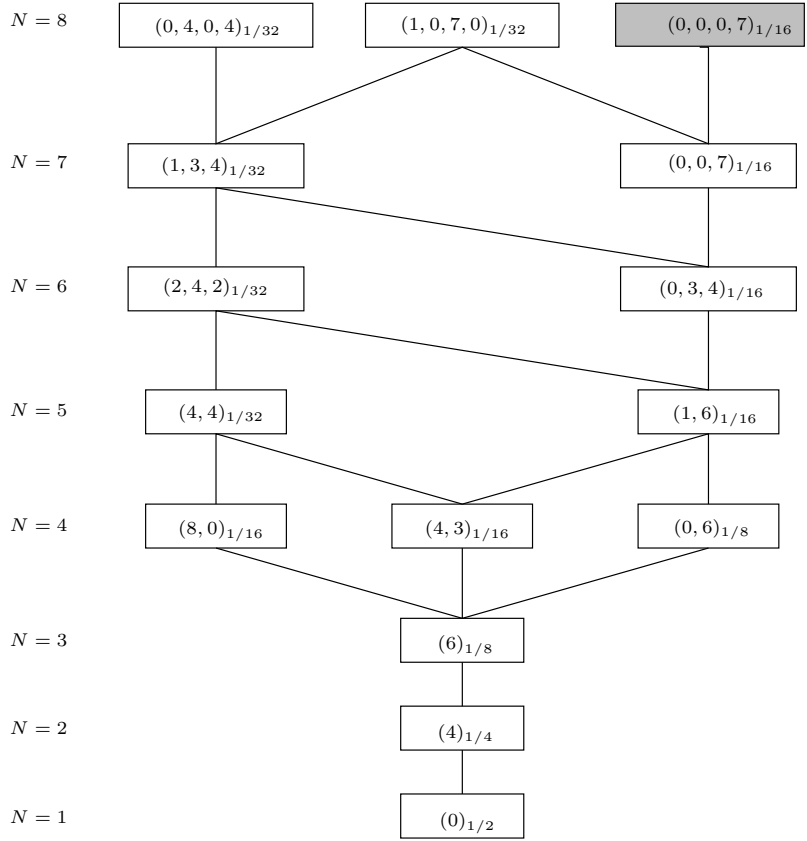}
\bigskip
\caption{{\bf $D$-brane intersections with $n=4$ in 10 dimensions:} {\it the
 numbers $(n_1, n_2,\ldots)$ label the number of times a building
 block with $(1,2,\ldots)$ worldvolume directions is used. The
 subscript in the Figure indicates the amount of supersymmetry
 preserved in each solution. The number $N$ indicates
the number of independent harmonics. The lines between solutions indicate
 how one configuration follows from another by adding (or deleting)
 a harmonic function. The configuration (0,0,0,7) cannot be
 extended to 11 dimensions in terms of (non-boosted) 2- and 5-branes only.}}
\label{D-tree}
\end{figure}
\bigskip

Equation (\ref{proj}) defines a projection operator on $\epsilon$
 that breaks half of the supersymmetry. For the IIA cases ($p$
 even) we have $\epsilon_{(p)}^\prime = \epsilon$
 for $p=0,4,8$, $\epsilon_{(p)}^\prime = \gamma_{11}\epsilon$
 for $p=2,6$; for IIB ($p$ odd) $\epsilon_{(p)}^\prime = i\epsilon$
 for $p=-1,3,7$, $\epsilon_{(p)}^\prime = i\epsilon^\star$
 for $p=1,5$, while always $\epsilon = H^{-1/8}\epsilon_0$, for
 constant $\epsilon_0$. It is known \cite{Pol2,BeBeJa} that a
 configuration of two intersecting $D$-branes can only be
 supersymmetric (keeping 1/4th of maximal supersymmetry) if they
 intersect in such a way that there are $n=4$ or 8 relative
 transverse directions, exactly the condition necessary to be also
 a solution of the equations of motion.

Adding more $D$-branes to the composite system, implies that more
 projection operators are introduced. Each time we add a new
 projection operator, half of the remaining supersymmetry gets
 broken. However, sometimes it is possible to add a $D$-brane in
 such a way that its projection operator is not independent,
 but given by a product of previous operators\footnote{A similar mechanism 
has been observed in lower dimensions, where $p$-brane solutions with 
different numbers of participating field strengths preserve the same amount 
of supersymmetry \cite{LuPo}.} \cite{Green,Klebanov,Gaunt}. In that
 case no additional supersymmetry generator is broken. In
 Figure (\ref{D-tree}) we see this happen for example in the $N=4$
 intersection. For $N=3$ we have one 0-brane and two  4-branes
 which preserve $1/8$th of the supersymmetry because
 of the three independent projection
operators
\begin{eqnarray}
&&(1 + \gamma_{0})\epsilon = 0\, , \nonumber \\
\label{proj-op}
&&(1 + \gamma_{01234})\epsilon = 0\, ,  \\
&&(1+ \gamma_{01256})\epsilon = 0\, . \nonumber
\end{eqnarray}
From Figure (\ref{D-tree}) we see that there are three different
 ways to add a fourth brane. Two of them break an extra half of
 the remaining supersymmetry (configurations (8,0) and (4,3)),
 since in these cases the new brane introduces an independent projection 
operator. The third way (corresponding to configuration (0,6))
is by adding a 4-brane oriented in such a
 way that its projection operator
\begin{equation}
(1 +  \gamma_{03456})\epsilon = 0\, 
\end{equation} 
is exactly the product of the previous three
 operators (\ref{proj-op}). In this way no extra conditions on
 the Killing spinor arise. 

The construction of projection operators for supersymmetry is
 another way of building up Figure (\ref{D-tree}).
 Apparently supersymmetry and the equations of motion
 go hand in hand: supersymmetry protects the stability of a
 configuration and vice versa, all stable
 solutions are supersymmetric\footnote{The procedure for constructing
all independent projection operators for ten-dimensional supersymmetry
resembles the procedure for constructing the
independent central charges of a supersymmetry algebra in lower than
ten dimensions (K.~Stelle, private communication).}. 
The amount of unbroken supersymmetry
 of each configuration can be found in Figure (\ref{D-tree}).

One $N=8$ solution in Figure ({\ref{D-tree}) is special. By
 using $T$-duality it cannot be expressed in terms of $2$-
 and $4$-branes only, and therefore it cannot be lifted to eleven
 dimensions as an intersection of (non-boosted) $M$-branes.
 It is the solution $(0,0,0,7)$, indicated by a grey
 box in Figure (\ref{D-tree}). We will discuss this solution in the next
 Section.

\section{Intersections of M-branes in 11 dimensions}

\noindent
The basic solutions in $D=11$ are the $M2$-brane
 solution  \cite{Duff-Stelle}: 
\begin{eqnarray}
 ds_{E,11}^2 &=& {H_2^{-2/3}}\  ds^2_3 - 
 {H_2^{1/3}} \ ds^2_8, \nonumber \\
 F_{012i}&=& \partial_i H_2^{-1}, 
\end{eqnarray}
\noindent
and the $M5$-brane solution\footnote{Note that we 
represent the fivebrane in terms of 
 a six-form gauge field, the field strength $F_{012345i}$ being the dual of
 $F_{jklm}$. This will be done for all $D=11$ solutions presented in 
 the paper. The $D=11$ Chern-Simons term does not contribute to
 the solutions considered in this paper. This can be easily seen
by noting that all nonzero gauge-field curvatures have a
time component.}
 \cite{Guven}:
\begin{eqnarray}
ds_{E,11}^2 &=& {H_5^{-1/3}}\  ds^2_6 - 
 {H_5^{2/3}} \ ds^2_5, \nonumber \\
 F_{012345i}&=& \partial_i H_5^{-1}, 
\end{eqnarray}
\noindent
where $H_2$ and $H_5$ are harmonic functions on the eight and five
 dimensional space transverse to the brane, respectively. As in the previous
 Section, we will construct all multiple intersections satisfying
 condition (1) in the Introduction, and obtain their residual
 supersymmetry. As stated in the Introduction, each pair
 of $M$-branes in a composite configuration must
 be $(0|2,2)$, $(3|5,5)$ ($n=4$)
 or $(1|2,5)$ ($n=5$) \cite{Papad-Town,Tseyt1}. 

We use the same representation as in the previous chapter: $\x$ for
 the world volume coordinates and $-$ for the transverse
 coordinates. The three allowed $N=2$ intersections of $M$-branes
 can therefore be represented as\footnote{Note that we cannot apply
$T$-duality in eleven dimensions to relate these three intersections. 
Therefore we must also consider intersections
of $M$-branes that intersect over a $p$-brane with $p > 0$.}:

\begin{equation} (0|2,2) : \ \ \ \left\{
\begin{array}{c|cc|cc|cc|cc|cc|}
\x& \x& \x&  -&  -&  -&  -&  -&  -&  -& - \\
\x&  -&  -& \x& \x&  -&  -&  -&  -&  -& -
\end{array}\right.
\end{equation}

\begin{equation} (1|2,5) : \ \ \ \left\{
\begin{array}{c|cc|cc|cc|cc|cc|}
\x& \x& \x&  -&  -&  -&  -&  -&  -&  -& - \\
\x& \x&  -& \x& \x& \x&  \x&  -&  -&  -& -
\end{array}\right.
\end{equation}

\begin{equation} (3|5,5) : \ \ \ \left\{
\begin{array}{c|cc|cc|cc|cc|cc|}
\x& \x& \x& \x& \x& \x&  -&  -&  -&  -& - \\
\x& \x& \x& \x&  -&  -& \x& \x&  -&  -& -
\end{array}\right.
\end{equation}

Next, we add further $M2$-branes and/or $M5$-branes, always satisfying 
 condition (1) for each pair. Like in $D=10$, we find that this procedure
stops at $N=8$. We will not present the details of our constructive
procedure but instead present the results below.
 It might be thought that the $D=11$ result
 can be immediately obtained from the results of Section 2, but
 this is not quite true. One can go from $M$-branes in $D=11$
 to $D$-branes in $D=10$ only if there is a direction such that all
 $M2$-branes are reduced to $D2$-branes, and all $M5$-branes
 to $D4$-branes. This will not be true in general, some configurations
 (which have $N\ge 4$)
 in $D=11$ will only reduce to $D=10$ intersections that involve 
NS-NS branes. 
 
To characterize the configurations, we use
 again the contents of the columns (the components of 
 the metric corresponding to the spacelike directions except the
overall transverse coordinates)
 in the representation of the
 metric. For an $N$-intersection each
 column can have $1,\ldots,N$ $\x$'s, indicating woldvolume directions.
 The numbers of columns with $k$
 worldvolume directions label the solutions, in the notation
 $\{n_1,\ldots,n_N\}$ (using curly brackets).
It is convenient to classify, in a first stage, the eleven-dimensional
intersections up to $T$-duality. $T$-duality works as follows in $D=11$ 
\cite{BeHuOr}.
Two $D=11$ solutions are called $T$-dual if, upon reduction
to $D=10$ dimensions they lead to $T$-dual $D$-brane configurations.
 These $T$-dual $D=11$ solutions can 
 be represented by the labels $(n_1,\ldots,n_{[N/2]})$ (using round
brackets) which were used
in the previous Section to label $T$-dual $D$-brane configurations. 
Such a classification in terms of $D=10$ solutions was also
 used by \cite{Papad-Town2} for $N=2$. Of course, this notation can
only be used for $D=11$ intersections that can be reduced to $D$-branes only.

The results we find in $D=11$ can be represented in three different ways. 
First of all, in Figure (\ref{M-tree})
we present the solutions up to $T$-duality in $D=11$. 
For those $M$-brane intersections that reduce to one of the $D$-brane 
intersections given in Figure (1), we use the same notation
$(n_1, \cdots ,n_{[N/2]})$ as in the previous Section. 
The gray rectangles indicate the solutions which
 necessarily contain NS-NS-branes in $D=10$, and for those the $D=11$ 
 notation $\{n_1, \cdots ,n_N\}$ is used. 
As in $D=10$, we can have at most eight intersecting
 branes. Secondly, in Table (1) we provide more 
 details about the contents of Figure (\ref{M-tree}) by showing all
 $D=11$ solutions that correspond to the same $D=10$ $D$-brane intersection.
 Finally, in Appendix B the $N=8$ intersections are given explicitly.
We have checked that these intersections indeed
solve the equations of motion.

As in $D=10$, the complete structure of the $D=11$ intersections can
 be recovered by the requirement of partially unbroken supersymmetry.
 Since the procedure is identical to the one used in $D=10$ we will
 not give the details. The amount of unbroken supersymmetry for the 
 different solutions is indicated in Figure (\ref{M-tree}).

As an example consider the intersection of
 seven $M$-5-branes: $\{0,0,7,0,0,0,2\}$. This solution has recently
been considered in \cite {Lavri}.
 This configuration cannot be extended
 to $N=8$ by adding another $M$-brane but is equivalent, via $T$-duality 
 in $D=10$, to a second $N=7$ configuration $\{0,0,6,2,0,0,0\}$ (see 
Table (1)).
 This $T$-dual $N=7$ configuration can be extended to $N=8$ 
as indicated in Figure (2). Note that the
third $N=7$ configuration $\{0,0,0,7,0,0,1\}$, belonging
 to the same $(0,0,7)$ class, and its extension to $N=8$
 $\{1,0,0,7,0,0,0,1\}$ were given in \cite{Gaunt}.

 Finally, we consider the 11-dimensional origin of the $N=8$ $D$-brane
 intersection  $(n_1,n_2,n_3,n_4)=(0,0,0,7)$. In Section 2 we found that 
 this intersection does not follow from  the
 dimensional reduction of a $D=11$ intersection consisting of 
 (non-boosted) 2- and 5-branes only.
 Instead we find that it corresponds to a 
 $N=7$ intersection $\{0,0,0,7,0,0,1\}$ boosted along
 the common string direction. This configuration can be viewed as an 
intersection of seven $M5$-branes and a $D=11$ gravitational wave
 and has a non-diagonal metric: 

\newpage
\begin{figure}[ht]
\centering
\input{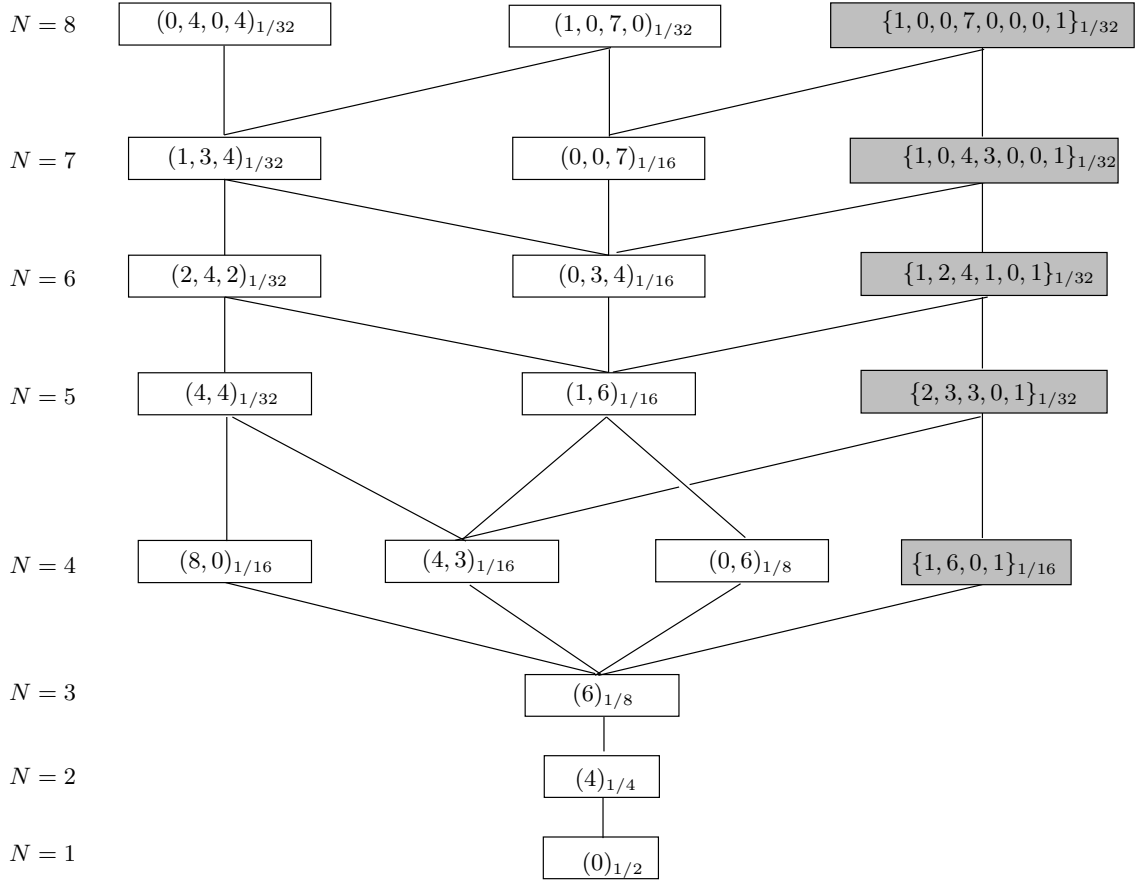}
\bigskip
\caption{{\bf $M$-brane intersections with $n=4,5$ in 11 dimensions:} 
{\it the numbers $(n_1, \cdots ,n_{[N/2]})$ are the same labels used in 
$D=10$, 
 and indicate to which $D$-brane intersection the $D=11$ solution
 reduces. 
 The configurations in gray rectangles only reduce to $D=10$ intersections
 involving NS-NS branes. For these configurations we use the eleven-dimensional
 notation $\{n_1,\cdots ,n_N\}$ explained in the text. 
 The subscripts indicate the amount of residual 
 supersymmetry.}}
\label{M-tree}
\end{figure}
\bigskip

\begin{eqnarray}
\label{wave1}
ds_{11}^2 = (H_2 H_3 H_4 H_5 H_6 H_7 H_8)^{- {1 \over 3}}
 [ (2-H_1) dt^2 - H_1dx_{10}^2 + 2(1-H_1)dtdx_{10}\nonumber\\
- (H_2H_5H_8)dx_1^2 - (H_2H_6H_7)dx_2^2 
 - (H_3H_6H_8)dx_3^2 - (H_3H_5H_7)dx_4^2\\
 - (H_4H_7H_8)dx_5^2  - (H_4H_5H_6)dx_6^2 
 - (H_2H_3H_4)dx_7^2 \nonumber\\
- (H_2H_3H_4H_5H_6H_7H_8)(dx_8^2 + dx_9^2) 
] \nonumber \, .
\end{eqnarray}
 The solution has
 two overall transverse directions, and all $H_i$ are harmonic on this
 two-dimensional space.

\newpage
{\scriptsize
\begin{tabular}{|p{10mm}|p{40mm}|p{40mm}|p{40mm}|}

\hline
\bf{N=8} & \bf{(0,4,0,4)} & \bf{(1,0,7,0)} &\bf{\{1,0,0,7,0,0,0,1\}}   \\
\hline
&[$2^4$,$5^4$]\it{\{0,4,0,5,0,0,0,0\}}&[$2^4$,$5^4$]\it{\{1,0,6,1,1,0,0,0\}}
&[$2^1$,$5^7$]\it{\{1,0,0,7,0,0,0,1\}} \\
\hline

\end{tabular}

\begin{tabular}{|p{10mm}|p{40mm}|p{40mm}|p{40mm}|}

\hline
\bf{N=7} & \bf{(1,3,4)} & \bf{(0,0,7)} &\bf{\{1,0,4,3,0,0,1\}} \\
\hline
& [$5^7$]\it{\{1,0,4,0,3,0,1\}} & [$5^7$]\it{\{0,0,7,0,0,0,2\}} & 
[$2^1$,$5^6$]\it{\{1,0,4,3,0,0,1\}} \\
&[$5^7$] \it{\{0,3,0,4,0,1,1\}} & [$5^7$]\it{\{0,0,0,7,0,0,1\}} &\\
&[$2^3$,$5^4$]\it{\{1,2,4,1,1,0,0\}} &[$2^3$,$5^4$]\it{\{0,0,6,2,0,0,0\}}&\\
&[$2^3$,$5^4$]\it{\{1,3,1,4,0,0,0\}}&			&\\
&[$2^4$,$5^3$]\it{\{1,3,4,1,0,0,0\}}&  			&\\
\hline
\end{tabular}

\begin{tabular}{|p{10mm}|p{40mm}|p{40mm}|p{40mm}|}
\hline
\bf{N=6} & \bf{(2,4,2)}		& \bf{(0,3,4)}	&\bf{\{1,2,4,1,0,1\}}\\
\hline
&[$5^6$]\it{\{1,2,2,2,1,1\}}&[$5^6$]\it{\{0,0,4,3,0,1\}}&
[$2^1$,$5^5$]\it{\{1,2,4,1,0,1\}}\\
&[$2^2$,$5^4$]\it{\{1,4,2,1,1,0\}} &[$5^6$]\it{\{0,3,4,0,0,2\}}	& \\
&[$2^2$,$5^4$]\it{\{2,2,2,3,0,0\}} &[$2^2$,$5^4$]\it{\{0,2,4,2,0,0\}}	& \\
&[$2^3$,$5^3$]\it{\{2,3,3,1,0,0\}} &[$2^3$,$5^3$]\it{\{0,3,5,0,0,0\}}	& \\
&[$2^4$,$5^2$]\it{\{2,5,2,0,0,0\}} & & \\
\hline
\end{tabular}

\begin{tabular}{|p{10mm}|p{40mm}|p{40mm}|p{40mm}|}
\hline
\bf{N=5} & \bf{(4,4)}	& \bf{(1,6)}        &\bf{\{2,3,3,0,1\}}\\
\hline
&[$5^5$]\it{\{2,2,2,2,1\}}&[$5^5$]\it{\{1,4,2,0,2\}}
   &[$2^1$,$5^4$]\it{\{2,3,3,0,1\}}\\
&[$2^1$,$5^4$]\it{\{3,1,3,2,0\}}&[$5^5$]\it{\{0,2,4,1,1\}}&\\
&[$2^2$,$5^3$]\it{\{3,3,2,1,0\}}&[$2^1$,$5^4$]\it{\{0,4,2,2,0\}}&\\
&[$2^3$,$5^2$]\it{\{4,3,2,0,0\}}&[$2^1$,$5^4$]\it{\{1,6,0,1,1\}}&\\
&[$2^4$,$5^1$]\it{\{5,4,0,0,0\}}&[$2^2$,$5^3$]\it{\{1,3,4,0,0\}}&\\
& &[$2^3$,$5^2$]\it{\{1,6,1,0,0\}}&\\
\hline
\end{tabular}

\begin{tabular}{|p{10mm}|p{28.95mm}|p{28.95mm}|p{28.95mm}|p{28.95mm}|}
\hline
\bf{N=4} &	\bf{ (8,0)}	& \bf{(4,3)}&\bf{(0,6)}&\bf{\{1,6,0,1\}}\\
\hline
&[$2^2$,$5^2$]\it{\{6,1,2,0\}}	&[$5^4$]\it{\{3,3,1,2\}} 
&[$2^2$,$5^2$]\it{\{0,7,0,0\}}    &[$2^1$,$5^3$]\it{\{1,6,0,1\}} \\
&[$2^4$]\it{\{8,0,0,0\}} &[$5^4$]\it{\{1,3,3,1\}} &[$5^4$]\it{\{0,6,0,2\}} & \\
&[$5^4$]\it{\{4,0,4,1\}} &[$2^1$,$5^3$]\it{\{4,3,1,1\}} & & \\
&		&[$2^1$,$5^3$]\it{\{2,3,3,0\}} & & \\
&		&[$2^2$,$5^2$]\it{\{3,4,1,0\}} & & \\
& 		&[$2^3$,$5^1$]\it{\{5,3,0,0\}} & & \\
\hline
\end{tabular}

\begin{tabular}{|p{10mm}|p{40mm}p{40mm}p{40mm}|}

\hline
\bf{N=3}	&\bf{(6)}&&	\\
\hline
&[$5^3$]\it{\{6,0,3\}}&[$5^3$]\it{\{3,3,2\}}
        &[$2^1$,$5^2$]\it{\{2,5,0\}} \\ 
&[$5^3$]\it{\{0,6,1\}}&[$2^1$,$5^2$]\it{\{5,2,1\}}
          &[$2^2$,$5^1$]\it{\{5,2,0\}} \\
&		&	 	&[$2^3$]\it{\{6,0,0\}} \\
\hline
\end{tabular}

\begin{tabular}{|p{10mm}|p{40mm}p{40mm}p{40mm}|}
\hline
\bf{N=2}	&\bf{(4)}&&	\\
\hline
&[$5^2$]\it{\{4,3\}}
&[$2^1$,$5^1$]\it{\{5,1\}}
&[$2^2$]\it{\{4,0\}} \\
\hline
\end{tabular}
\bigskip
\begin{table}[h]
\label{Mtable}
\caption{{\bf Table of $M$-brane intersections in D=11:}
 {\it The number $N$ indicates the number of independent harmonics.
The boldface labels $(n_1,\ldots, n_{[N/2]})$ 
correspond to the $D=10$ $D$-brane
 intersection to which the $D=11$ solutions reduce (when applicable).
The numbers between
square brackets indicate the number of $M2$-branes and $M5$-branes involved
in the intersection.
The labels $\{n_1,\ldots ,n_N\}$ specify the structure of the
$D=11$ metric as explained in the text. 
}} 
\end{table}
}
\bigskip

\section{Reduction to lower dimensions}

A natural application of our results is the reduction of the $M$-brane and 
$D$-brane intersections we found in the previous two Sections
to dilatonic $p$-branes in lower dimensions. This 
will lead to dilatonic $p$-brane solutions which can be understood as 
$D$- and/or $M$-brane bound states in $D=10,11$. The interpretation of 
lower dimensional solutions in terms of bound states of $D$- and/or $M$-branes
in $D=10,11$ is a useful tool for understanding the properties of these lower
dimensional solutions, especially in the case of (extremal) black holes
where it has opened up the possibility for a microscopic explanation
of the Bekenstein-Hawking entropy \cite{StV}.
It was recently discovered that the $D=4$ (extremal) dilaton black 
holes preserving 
$\tfrac{1}{2}$ of the supersymmetry can be interpreted as bound 
states of
$D$-branes ($M$-branes) compactified on a six-torus (seven-torus)
\cite{Papad-Town,Tseyt1,Klebanov,Gaunt,BaLa,BeBe}. It was shown, using
the $N=4$ (0,6) intersection in $D=10$ (see Figure (1)), 
that the four values of the 
dilaton coupling $a^2$ in $D=4$ could be reproduced by
identifying the harmonic functions (equal charges) and truncating to 
intersections with smaller $N$ (by setting some of the harmonic functions
 equal to one).

Using the $D$- and $M$-intersections  constructed in this paper
we find many other intersections
which can  be reduced to $p$-branes in lower dimensions. The 
general (Einstein frame) form of our reduced action (upon identifying some
of the  harmonic functions and setting the others equal to one) 
for $D>2$ will always be in the class of Lagrangians of the form
\be
{\cal{L}}_{E,D}=\sqrt{g} [R+\tfrac{1}{2} (\pd \phi)^2 + 
 { (-1)^{p+1} \over 2(p+2)!} e^{a \phi} F_{(p+2)}^2].
\ee
Using the Ansatz
\bea
ds^2 _{E,D} &=&H^{\alpha} ds_{p+1}^2 -H^{\beta} ds^2 _{d-p-1} , \nn \\
\label{gensatz}
e^{2\phi} &=&H^{\gamma} , \\
F_{0..p i} &=&\delta \ \pd_{i} H^{-1} \nn ,
\eea
we know that the general $p$-brane solution $(D>2)$ is given by\footnote{
We use here a form of the solution as given in \cite{Lu}.}
\bea
\alpha &=&-{4(D-p-3)\over \Delta(D-2)} \qquad , \qquad 
\label{gensol}
\beta   =  {4 (p+1)\over \Delta(D-2)} , \nn \\
\gamma &=&{4a\over\Delta} \qquad , \qquad \delta^2={4\over \Delta}\,,
\eea
with 
\be
\Delta = a^2 + 2 {(p+1)(D-p-3)\over D-2}\,.
\label{Delta}
\ee
The lower dimensional
 $p$-brane solutions which follow from
 the reduced $D$-brane and $M$-intersections (now containing only
one independent harmonic function) must fall inside
 this class of solutions. For supersymmetric solutions
 one must have \cite{Lu,LuPo}
\be
\label{susa}
\Delta={4}/{N}\,,
\ee 
where $N$ is an integer labeling the number of 
 participating field strengths, in our case this is the number
 of intersecting branes. 

Any toroidal Kaluza-Klein
 reduction of the $D=10,11$ intersections 
 will be a supersymmetry preserving $p$-brane solution
 in a lower dimension. 
 Because the number of participating
 field strengths is equal to the number of intersecting branes we can 
 immediately read off our dilatonic $p$-brane solution from 
 (\ref{gensol}) and (\ref{Delta})
 
 As an illustration, consider the $N=8$ $D$-brane intersections (see
Figure (1)). We see
 that one of them, labeled by (0,0,0,7), can be naturally reduced to 
 $0$-branes in $D=3$ 
 by reducing over all relative transverse directions.
 Every truncation of this solution can of course also be 
 reduced to $D=3$ $0$-branes, giving rise to 8 different supersymmetry 
 preserving solutions in $D=3$.
 Doing the explicit Kaluza-Klein reduction we find that the different values 
 of $a^2$ representing the different solutions (the explicit solution can be 
 determined using (\ref{gensol})) are given by
\be 
  a^2=\frac{4}{N} ,
\label{p0d3}
\ee 
which is just (\ref{Delta}) with $p=0$, $D=3$ and $N$ running from 1 to 8. 
 So we find 8 supersymmetry preserving $0$-branes
 in $D=3$ (in contrast to the 4 $0$-branes in $D=4$) with the dilaton
 coupling given by (\ref{p0d3}) \cite{LuPoSte}.

The general rule is to find the highest $N$ intersection in the $D$- and/or
 $M$-intersections
  that can be reduced to a single $p$-brane in a lower dimension. 
The $p$-brane solutions in that lower dimension are given by 
 (\ref{gensol}) and (\ref{Delta}) with $\Delta = 4/N$.
 Note that $N$ is the only parameter, and that therefore different 
 configurations of
 intersecting $D$- and/or $M$-branes with the same $N$,
  will all reduce to the same 
 $p$-brane in lower dimensions
 upon identification of the harmonic functions
 (even if the $D=10,11$
 intersecting solutions preserve different amounts of 
 supersymmetry).

We will now discuss the various $p$-branes in lower dimensions obtained after
reduction of $D$- and/or $M$-brane intersections.

\begin{itemize}
\item $D=3$:

So far we discussed the reduction of 
$D$-brane intersections to $0$-branes in $D=3$. Because
the $N=8$ configuration (0,0,0,7) cannot be oxidated to a (non-boosted)
 $D=11$
$M$-brane intersection, the reduction from $D=11$ to $0$-branes in $D=3$ will 
give only 7 different solutions. To be precise, the $N=7$ $M$-brane 
intersections labeled by $\{0,0,6,2,0,0,0\}$ and 
$\{0,0,0,7,0,0,1\}$ can be reduced 
to $D=3$ 0-branes (giving the same solutions as reduction of the $N=7$ 
(0,0,7) $D$-brane intersection). Of course, if we extend our Ansatz and use
the boosted $D=11$ $N=7$ $\{0,0,0,7,0,0,1\}$ solution (\ref{wave1})
we do obtain the 8th 
0-brane in $D=3$. 

In $D=3$ the other possibility is to consider 
 string (domain wall\footnote{We use the name domain wall to indicate a
$(D-2)$-brane solution in $D$ dimensions.}) solutions. Because we have 8 0-branes in $D=3$ and
 we can always do $T$-duality in one of the overall transverse directions on
 the $N=8$ (0,0,0,7) $D$-brane intersections we find 8 domain wall solutions.
 These same domain wall solutions also have an 
 $M$-brane bound state interpretation because the $\{1,0,0,7,0,0,0,1\}$
 $M$-brane
 intersection can be reduced to domain walls in $D=3$ (note that this
 intersection cannot be reduced to $D=10$ $D$-branes). 

This means that in
 $D=3$ we find 8 0-branes and 8 domain walls, both series of solutions can 
 be described as bound states of $D$-branes. One of the 0-branes  
 cannot be interpreted as a bound state of (non-boosted) $M$-branes, 
 but can be obtained from 
 $D=11$ using a boosted $N=7$ solution.

\item $D=4$

Having a closer look at the $\{0,0,0,7,0,0,1\}$ $M$-brane 
intersection we see that this intersection can be reduced to strings 
(1-branes) in $D=4$, which implies that
there are 7 supersymmetry preserving string 
solutions in $D=4$ \cite{LuPo} with the solutions given by  
(\ref{gensol}) and (\ref{Delta}). From the reduction of the $D$-intersections
we can  obtain
4 different string solutions in $D=4$ by using $T$-duality in one of
the overall transverse directions in $D=10$ on the (0,6) intersection. So
we get 3 extra string solutions in $D=4$ from the $M$-brane intersections.

By performing $T$-duality on an overall transverse direction we obtain
4 domain walls coming from the (0,6) $D$-brane intersection \cite{BeRoPa}. 
Surprisingly,
we can get 3 extra domain-wall solutions in $D=4$ from the $M$-intersections.
 The $\{0,0,7,0,0,0,2\}$ $N=7$ $M$-brane intersection can be  
reduced to domain walls in $D=4$ and will give three extra 
domain wall solutions \cite{Lavri}. 

This completes the $D=4$ case, which has 4 0-branes,
7 strings  and 7 domain walls (4 coming from $D$-branes, 7 coming from 
$M$-branes).

\item $D>4$

In dimensions higher than four there are fewer possibilities. In $D=5$ we 
find three particle, three string, three membrane and three domain wall 
solutions coming from the $\{6,0,0\}$, 
 $\{0,6,1\}$, $\{3,3,2\}$ and $\{6,0,3\}$ $M$-brane 
intersections respectively \cite{Tseyt1}. Only the $N=1,2$ have a $D$-brane 
origin, all the solutions have an $M$-brane origin. 

In $D=6$ $p$-branes come 
in pairs and have a $D$- and $M$-brane origin. In $D=7$ there exist two
supersymmetry preserving 0-branes, both having an $M$-brane interpretation, 
only one having a $D$-brane interpretation. The basic $D$-branes 
in $D=10$ and/or the basic $M$-branes in $D=11$ ($M2$- and $M5$-brane) 
can be reduced to supersymmetry preserving $p$-branes in $D>7$. Because 
the $D8$-brane in $D=10$ has no (known) $D=11$ origin there will be no 
domain wall solution in $D=9$ with an $M$-brane origin. 

\item $D=2$

So far we did not discuss $D=2$. We see
that in principle all $N=8$ intersections in $D=10,11$ can be 
reduced to $D=2$ 0-branes. In this case, however, we must work
in the string frame and
(\ref{Delta}) is no longer valid. Therefore we redo the
Kaluza-Klein reduction of the $D=10$ intersections, keeping the 
string-frame metric. The reduction to $D=2$ 
will always fall in the following class of Lagrangians (only 0-branes)
\be 
{\cal{L}}_{S,2}=\sqrt{|g|}e^{-2\phi} [R-4(\pd \phi)^2] 
 -\tfrac{1}{4} \sqrt{|g|}  e^{a\phi} F_{(2)}^2 .
\label{lag2}
\ee
Weyl invariance in $D=2$ ensures that the reduced Lagrangian can always be
written in the above way.
The general 0-brane solution of this Lagrangian is
\bea
\alpha &=& \frac{2}{a} -1 \qquad , \qquad \beta=-\frac{2}{a} -1 ,\\
\gamma &=& \frac{2}{a} \qquad , \qquad \delta^2= -\frac{4}{a} ,
\eea
where it is understood that the same Ansatz (\ref{gensatz}) is used as 
in the Einstein frame case.

We 
find that for every number $N$ of intersecting $D$-branes there is only one 
dilaton coupling constant representing the 0-brane solution. For example, all 
three $N=8$ intersections give rise to the same two-dimensional 
$0$-brane, thus confirming that for every $N$ there is only one
0-brane solution, just like in $D>2$ \cite{Lu}. The dilaton couplings 
representing the supersymmetric solutions are given by
\be
a=-\frac{4}{N} ,
\label{susya2}
\ee 
with $N=1,\ldots,8$. Note that we now give the string frame dilaton
coupling (defined by the Lagrangian in (\ref{lag2})) with a definite 
sign. This is done because in the $D=2$ string frame there is no
 symmetry that flips the sign.
Because 
the different solutions are labeled by $N$ and most of the $M$-brane
intersections can be reduced to $D$-branes in $D=10$ we are convinced that the 
 reduction from $D=11$ will give the same results. 
So there are 8 supersymmetry 
preserving 0-brane solutions in $D=2$ with dilaton coupling given in 
(\ref{susya2}), all of them having a $D$- and $M$-brane interpretation.
It would be interesting to see whether
the $D$-brane interpretation could shed any new light on the
structure of black holes in two dimensions \cite{Witbh}\footnote{In fact, 
O.A.~Soloviev informed us that he is studying this connection.}.
\end{itemize}
\bigskip
\begin{table}[h]
\[
\begin{array}{||c|c|c|c|c|c|c|c||} \hline
&D& p=0        & p=1        & p=2        & p=3        & p=4       & \\ \hline
&6& 2[D^2,M^2] & 2[D^2,M^2] & 2[D^2,M^2] & 2[D^2,M^2] & 2[D^2,M^2]& \\ \hline
&5& 3[D^2,M^3] & 3[D^2,M^3] & 3[D^2,M^3] & 3[D^2,M^3] & -         & \\ \hline
&4& 4[D^4,M^4] & 7[D^4,M^7] & 7[D^4,M^7] & -          & -         & \\ \hline
&3& 8[D^8,M^7] & 8[D^8,M^8] & -          & -          & -         & \\ \hline
&2& 8[D^8,M^8] & -          & -          & -          & -         & \\ \hline 
\end{array} 
\]
\bigskip
\caption{{\bf $D$-brane and/or $M$-brane interpretation of dilatonic 
$p$-branes in $D\le 6$ dimensions:}
{\it The numbers $r[D^s, M^t]$ indicate that there are $r$ solutions,
for given $D,p$, of which $s$ have a D-brane interpretation and
$t$ have an (non-boosted) $M$-brane interpretation. Note that one of the 
eight $D=3$ $p=0$ branes has a $D=11$ interpretation only 
as a boosted $N=7$ 5-brane solution.}}
\label{lowerp}
\end{table}
\bigskip

Finally, we mention that all $p$-brane solutions in lower 
dimensions preserve half of the maximal supersymmetry in contrast to the 
intersecting $D$- and/or $M$-intersections in $D=10,11$. This gain in 
supersymmetry is a result of the identification of the different harmonics 
(equal charges). For an overview of the number of dilatonic $p$-brane solutions
in lower dimensions ($D\le 6$) with a $D$- and/or (non-boosted) 
$M$-brane bound state interpretation
we refer to Table (\ref{lowerp}).

\section{Conclusions}

In this paper we have given a classification of all multiple intersections
of $D$-branes in ten dimensions and $M$-branes in eleven dimensions that 
correspond to threshold BPS bound states. In both cases we found that
the maximum number of participating branes is eight. Allowing one $n=8$ pair
we can extend the number of intersecting branes to the maximum of nine.  
Furthermore, we found that not all $D$-brane intersections can be lifted up 
to non-boosted $M$-brane intersections in eleven dimensions. Conversely,
not all $M$-brane intersections can be reduced to a ten-dimensional
configuration of intersecting $D$-branes only. We also investigated the
supersymmetry of the intersections, both in ten and eleven dimensions,
and found that for all configurations at least 1/32 of the supersymmetry
is preserved.

There are several ways in which the classification presented in this work
can be extended. First of all, we may consider boosted $M$-brane 
intersections \cite{Papad-Town,Tseyt1,Klebanov,Gaunt,Khviengia, Costa}. 
The rule seems to be
that, in case the intersection has a common string isometry direction $x$,
one can add a Brinkmann wave with nontrivial (non-diagonal)
metric components in the $(x^0,x)$ direction. The Brinkmann wave is the
11-dimensional origin of the $D0$-brane. A similar mechanism
should exist where the wave is replaced by its magnetically charged
partner (being the 11-dimensional origin of the $D6$-brane).
We thus obtain intersections 
with more than eight independent harmonics. It is expected
that this wider class of intersecting $M$-branes gives rise, upon
dimensional reduction to ten dimensions, to the class of intersections
that contains not only $D$-branes but also NS-NS strings, five-branes
and/or their $T$-duals. Of course, these intersections do not involve 
8-branes whose 11-dimensional origin sofar has been a mystery.
This concludes the classification of solutions that correspond to
threshold BPS bound states.

One may also extend the solutions to the ones that correspond  to
non-threshold BPS bound states. For instance, by considering $M$-branes
finitely boosted in a transverse direction \cite{Tseyt3}
one obtains $D=11$
solutions that reduce to $D$-brane bound states with $n=2$ or $6$. Furthermore,
by considering a wave
propagating along a generic cycle of a 2-torus \cite{Tseyt3} one obtains
$D=11$ solutions that reduce to the $D=10 (q_1,q_2)$ string solutions of
\cite{Schwarz}. There are also non-threshold BPS bound states in 
eleven dimensions, like the one given in \cite{To}. It would be interesting to
see how they fit in the general classification scheme.

Finally, we note that all knowledge about intersecting configurations
is contained in the $D=11$ solution with the maximum number of 
independent harmonics.
The other ones can be obtained from these basic solutions via truncation
and/or dimensional reduction.
We have seen that there are very few of these basic configurations.
Even including intersections with NS-NS branes and/or non-threshold
BPS bound states, we expect the number of basic solutions to
be limited. It would be of interest to construct these basic $M$-brane
configurations explicitly.

\vspace{1truecm}
\noindent{\bf Acknowledgements}
\bigskip

\noindent
We thank A.~Ach\'ucarro, A.~Hams, J.~Nijhof and P.~Townsend 
 for useful discussions. 
 This work is part of the research 
 program of the ``Stichting
 voor Fundamenteel Onderzoek der Materie'' (FOM). 
 It is also supported  by the European Commission TMR programme 
 ERBFMRX-CT96-0045,
 in which E.B. and M.~de R. are associated to the University of Utrecht.
\newpage

\appendix

\section{$N=8$ $D$-brane intersections}
\label{N=8}
In this Appendix we give the explicit form of the metric for 8 intersecting 
$D$-branes. There are three inequivalent $N=8$ intersections.
Note that the first two can be written via $T$-duality (in the $x_2$ and $x_3$
direction) as an 
intersection of four 2-branes and four 4-branes, and therefore can be 
lifted up to intersecting (non-boosted)
$M$-brane solutions in eleven dimensions. The third
solution can not be written as intersecting 2- and 4-branes and 
requires a non-diagonal form of the metric in eleven dimensions 
(see eq.~(\ref{wave1})). Using the notation explained in Section 2, the
metric for the three $N=8$ intersections are given by

{\scriptsize
\begin{equation}   (0,4,0,4) : \ \ \ \left\{
\begin{array}{c|cc|cc|cc|cc|c}
\x & - & - & - & - & - & - & - & - & -  \\
\x & \x& \x& \x& \x& - & - & - & - & -  \\
\x & \x& \x& - & - & \x& \x& - & - & -  \\
\x & - & - & \x& \x& \x& \x& - & - & -  \\
\x & \x& - & \x& - & \x& - & \x& - & -  \\
\x & - & \x& \x& - & \x& - & - & \x& -  \\
\x & \x& - & \x& - & - & \x& - & \x& -  \\
\x & - & \x& \x& - & - & \x& \x& - & -
\end{array} \right.
\label{N=8a}
\end{equation}

\begin{equation}   (1,0,7,0) : \ \ \ \left\{
\begin{array}{c|cc|cc|cc|cc|c}
\x & - & - & - & - & - & - & - & - & -  \\
\x & \x& \x& \x& \x& - & - & - & - & -  \\
\x & \x& \x& - & - & \x& \x& - & - & -  \\
\x & - & - & \x& \x& \x& \x& - & - & -  \\
\x & \x& - & \x& - & \x& - & \x& - & -  \\
\x & - & \x& \x& - & \x& - & - & \x& -  \\
\x & - & \x& - & \x& \x& - & \x& - & -  \\
\x & - & \x& \x& - & - & \x& \x& - & -
\end{array} \right.
\label{N=8b}
\end{equation}

\begin{equation}   (0,0,0,7) : \ \ \ \left\{
\begin{array}{c|cc|cc|cc|cc|c}
\x & - & - & - & - & - & - & - & - & -  \\
\x & \x& \x& \x& \x& - & - & - & - & -  \\
\x & \x& \x& - & - & \x& \x& - & - & -  \\
\x & - & - & \x& \x& \x& \x& - & - & -  \\
\x & \x& - & \x& - & \x& - & \x& - & -  \\
\x & \x& - & - & \x& - & \x& \x& - & -  \\
\x & - & \x& \x& - & - & \x& \x& - & -  \\
\x & - & \x& - & \x& \x& - & \x& - & -
\end{array} \right.
\label{lonelyboy}
\end{equation}}

All intersections with $N<8$ can be obtained from these via different 
truncations. There are, of course, many different ways of truncating to
lower intersections. However, as we can see in Figure (\ref{D-tree}),
many of them will lead to the same class. To recognize the class one has
to determine the $n_{k}$'s representing the class, this means counting the 
times a particular building block occurs in the configuration (see Section 2). 

\section{$N=8$ $M$-brane intersections}

In this Appendix we give the explicit form of the metric for the 
configurations with $N=8$ $n=4,5$ $M$-branes. As explained in Section 3 
we can label classes of $M$-brane intersections by the $D$-brane intersection
classes they reduce to. Some possible $M$-brane
intersections cannot be reduced to $D$-brane intersections and then we
are forced to use the particular $D=11$ building block numbers. The
configurations below are given with the $D=11$ building block numbers 
(and their preserved supersymmetry). The first one can also be labelled 
with $(0,4,0,4)_{1/32}$ representing the $N=8$ $D$-brane intersection it 
reduces to. The second one reduces to the $(1,0,7,0)_{1/32}$ $N=8$ $D$-brane
intersection and the third one cannot be reduced to a $D$-brane 
intersection.   

{\scriptsize
\begin{equation} \{0,4,0,5,0,0,0,0\}_{1/32} : \ \ \ \left\{
\begin{array}{c|cc|cc|cc|cc|cc}
\x & \x& \x& - & - & - & - & - & - & - & -  \\
\x & - & - & \x& \x& - & - & - & - & - & -  \\
\x & - & - & - & - & \x& \x& - & - & - & -  \\
\x & - & - & - & - & - & - & \x& \x& - & -  \\
\x & - & \x& \x& - & \x& - & \x& - & \x& -  \\
\x & \x& - & - & \x& \x& - & \x& - & \x& -  \\
\x & \x& - & \x& - & - & \x& \x& - & \x& -  \\
\x & \x& - & \x& - & \x& - & - & \x& \x& - 
\end{array} \right.
\end{equation}

\begin{equation} \{1,0,6,1,1,0,0,0\}_{1/32} : \ \ \ \left\{
\begin{array}{c|cc|cc|cc|cc|cc}
\x & \x& \x& - & - & - & - & - & - & - & -  \\
\x & - & - & \x& \x& - & - & - & - & - & -  \\
\x & - & - & - & - & \x& \x& - & - & - & -  \\
\x & - & - & - & - & - & - & \x& \x& - & -  \\
\x & - & \x& - & \x& \x& - & - & \x& \x& -  \\
\x & \x& - & - & \x& \x& - & \x& - & \x& -  \\
\x & - & \x& \x& - & \x& - & \x& - & \x& -  \\
\x & \x& - & \x& - & \x& - & - & \x& \x& - 
\end{array} \right.
\end{equation}

\begin{equation} \{1,0,0,7,0,0,0,1\}_{1/32} : \ \ \ \left\{
\begin{array}{c|cc|cc|cc|cc|cc}
\x & \x& \x& - & - & - & - & - & - & - & - \\  
\x & \x& - & \x& \x& \x& \x& - & - & - & - \\
\x & \x& - & \x& \x& - & - & \x& \x& - & - \\
\x & \x& - & - & - & \x& \x& \x& \x& - & - \\
\x & \x& - & \x& - & \x& - & \x& - & \x& - \\
\x & \x& - & - & \x& - & \x& \x& - & \x& - \\
\x & \x& - & - & \x& \x& - & - & \x& \x& - \\
\x & \x& - & \x& - & - & \x& - & \x& \x& -
\end{array} \right.
\end{equation}}

As in the case of $D$-branes we can obtain lower intersections 
through truncation. In order to obtain all configurations in 
Table~(1) we have to make use of $T$-duality in $D=10$. This 
$T$-duality should be carried out
such that we keep $D$2 and/or $D$4-branes in $D=10$, so we can
lift up the solution back to intersecting $M$-branes in $D=11$. 
The $D=11$ building
block numbers ($n_k$) can then be read off in a straightforward manner.

\section{$D$-brane and $M$-brane intersections with $n=4,5,8$ relative transverse directions}
\label{n=8}

In this Appendix we discuss intersections with $n=4,8 (D=10)$ or
$n=4,5,8 (D=11)$, 
and dependence on relative transverse coordinates, i.e. corresponding to
the conditions (2) and  (3) in the Introduction.

It is easy to see that it is impossible to construct a configuration
with three intersecting $D$-branes such that all pairs have $n=8$
with a nontrivial dependence of the harmonic functions on the
relatively transverse coordinates.
 
Let us work out in some detail how the dependence on relative coordinates
 can be brought in, since we will have to be careful about the
 allowed dependence on these coordinates.
 Consider any $n=4$ configuration with dependence on overall
 transverse coordinates only, e.g.~the $D=10$ solution 
given in (\ref{N=8a}), 
which we copy below in (\ref{CN=8a}).
 Now, suppose $H_2$, the harmonic function corresponding to the second line 
 in (\ref{CN=8a}), does not depend on $x_9$ but instead on
 the relative coordinates $x_5,\ldots,x_8$.
 Then we have realized a configuration satisfying
 condition (2). However, in verifying the equations of motion we
 find that the dependence on $x_5,\ldots,x_8$ has to be further
 restricted: the metric components $g_{ii}$ have to be the same
 for each of the relative coordinates $x_i$ on which the brane depends.
 Only then do the equations of motion lead to
 a harmonic equation for $H_2$.
 In this case that means that $H_2$ can depend on only one of the coordinates
 $x_5,x_6,x_7$ or $x_8$.
 Note that we can do this only for one harmonic function at the time,
 since any pair which both depend on relative coordinates must have 
 $n=8$.

{\scriptsize
\begin{equation} N=8,\  (0,4,0,4) : \ \ \ \left\{
\begin{array}{c|cc|cc|cc|cc|c}
\x & - & - & - & - & - & - & - & - & -  \\
\x & \x& \x& \x& \x& - & - & - & - & -  \\
\x & \x& \x& - & - & \x& \x& - & - & -  \\
\x & - & - & \x& \x& \x& \x& - & - & -  \\
\x & \x& - & \x& - & \x& - & \x& - & -  \\
\x & - & \x& \x& - & \x& - & - & \x& -  \\
\x & \x& - & \x& - & - & \x& - & \x& -  \\
\x & - & \x& \x& - & - & \x& \x& - & -
\end{array} \right.
\label{CN=8a}
\end{equation}}

\noindent
Since we have a different dependence on one harmonic function, we also have 
to change our Ansatz for the gauge field: the gauge field of the $D$-brane
represented by $H_2$ is now given by 
\be
F_{01234r}=\partial_r H_2^{-1},
\ee
where $x_r$ can be either $x_5, x_6,x_7$ or $x_8$. 

It turns out
that in this way all configurations satisfying condition
(2) of the Introduction can be obtained from the ones satisfying
condition (1) of the Introduction. Therefore, the classification of
intersections satisfying condition (2) is the same as the 
one satisfying condition (1) and is therefore given by Figure (\ref{D-tree})
as well. This concludes our classification of the configurations
satisfying condition (2).

\newpage
\begin{figure}[ht]
\centering
\input{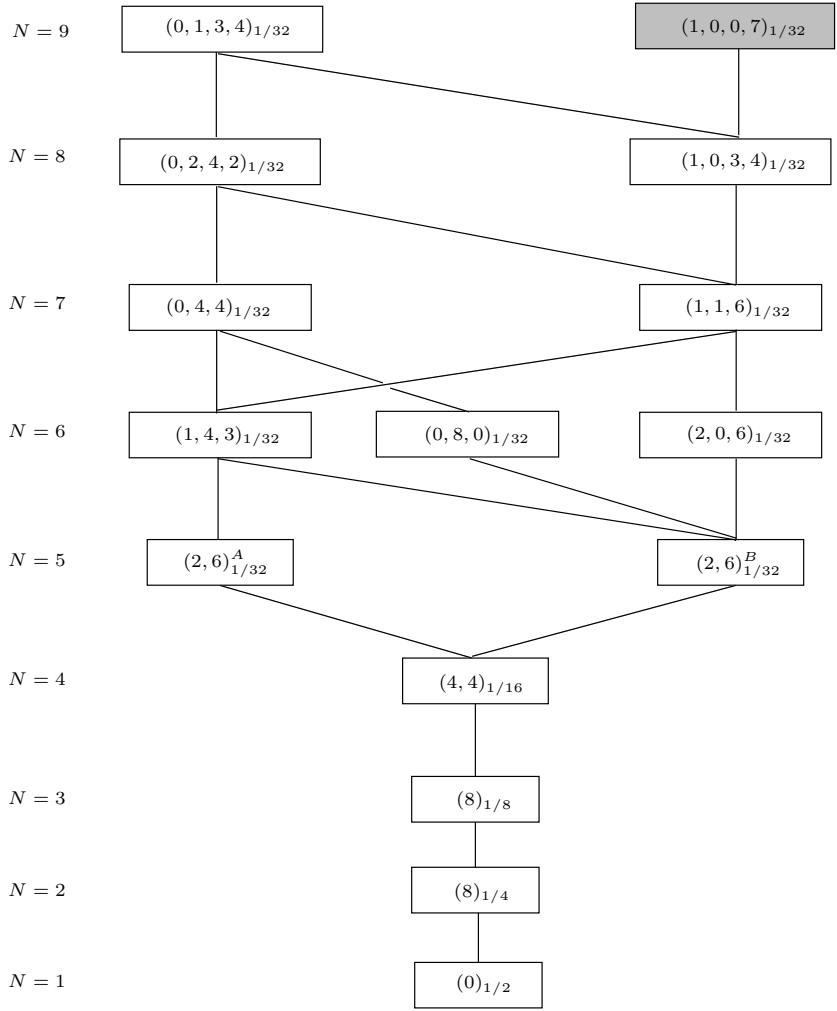}
\bigskip
\caption{{\bf $D$-brane intersections
  with $n=4,8$ in 10 dimensions:}\ {\it The  solutions are labelled by 
 $(n_1,\ldots n_{[N/2]})$, as explained in section 2. For $N=5$ an
extra superscript is added to distinguish between the two sets of labels.
 Subscripts indicate the supersymmetry of the 
 configurations. The $(1,0,0,7)$ configuration given in the
grey rectangle cannot be extended
 to eleven dimensions in terms of (non-boosted) 2- and 5-branes.}}
\label{D-tree4,8}
\end{figure}

We next consider the configurations satisfying condition (3) of the
Introduction.
Consider the ``mirror'' configuration of $H_2$ in the
above configuration, i.e.~the brane
\begin{equation} 
\begin{array}{c|cc|cc|cc|cc|c}
\x & - & - & - & - & \x & \x& \x & \x & - \,,
\end{array} 
\end{equation}
in which all $\x$'s in the relative coordinates have been replaced by $-$'s
 and vice versa. This 4-brane has $n=8$ with $H_2$, and $n=4$ with
 the other seven branes included in (\ref{CN=8a}). Since it has
 $n=8$ with $H_2$, its harmonic function, $H_9$ must depend on
 (some of the) coordinates $x_1,\ldots,x_4$, to satisfy the conditions (3)
of the Introduction. An investigation of the 
equations of motion reveals that only dependence on
 one of the coordinates  $x_1,x_2,x_3$ or $x_4$ is allowed: the metric 
must again be of the same form
in the relative transverse coordinates.
 
This simple mechanism makes it possible to introduce an additional brane
 into any $n=4$ configuration, by constructing an $n=8$ pair with one
 of the constituents. In the present case this leads to the $N=9$
 configuration:

{\scriptsize
\begin{equation} N=9,\  (0,1,2,5) : \ \ \ \left\{
\begin{array}{c|cc|cc|cc|cc|c}
\x & - & - & - & - & - & - & - & - & -  \\
\x & \x& \x& \x& \x& - & - & - & - & -  \\
\x & \x& \x& - & - & \x& \x& - & - & -  \\
\x & - & - & \x& \x& \x& \x& - & - & -  \\
\x & \x& - & \x& - & \x& - & \x& - & -  \\
\x & - & \x& \x& - & \x& - & - & \x& -  \\
\x & \x& - & \x& - & - & \x& - & \x& -  \\
\x & - & \x& \x& - & - & \x& \x& - & -  \\
\x & - & - & - & - & \x& \x& \x& \x& -  \\
\end{array} \right.
\label{N=9,10}
\end{equation}}

In Figure (\ref{D-tree4,8}) we give the extension to $n=4,8$ of 
 Figure (\ref{D-tree}). 
Note that, in contrast to the $n=4$ case, the labels $(n_1,\ldots,
n_{[N/2]})$ do not uniquely 
specify the configuration:
 the two configurations with $N=5$ have the same building block numbers 
($n_k$'s), although they are inequivalent. To distinguish between them we
have added a superscript $A$ or $B$. For the sake of completeness we
also give the form of the gauge fields of the $D$-branes that depend
on the relative coordinates $x_r$ and $x_s$ \cite{BeBeJa}
\be
F^{(2)}_{01234r} = H_9 \partial_r H_2^{-1}, \hspace{10 mm}
F^{(9)}_{05678s} = H_2 \partial_s H_9^{-1}.
\ee

Note that these curvatures indeed satisfy the Bianchi identity, and that we 
obtain the correct truncation by setting either $H_2$ or $H_9$ equal to one.
\bigskip

We next repeat this analysis for $M$-branes in $D=11$.
The only possibility for two $M$-branes to have $n=8$ is 
 two 5-branes intersecting over a string \cite{Gaunt}:

{\scriptsize
\begin{equation} n=8,\ (1|5,5) : \ \ \ \left\{
\begin{array}{c|cc|cc|cc|cc|cc}
\x& \x& \x& \x& \x& \x&  -&  -&  -&  -&  -\,  \\
\x&  -&  -&  -&  -& \x& \x& \x& \x& \x&  -\,  
\end{array}\right.
\end{equation}}
This solves the equations of motion with  $H_1$, $H_2$ 
 depending on the relative
 transverse directions. 
 Again an extension to $N=3$ is impossible, but we can follow
 the same construction as in $D=10$. Since there are no essential differences
 between the $D=10$ and the $D=11$ construction we will skip the details.
 Figure (\ref{8M-tree}) represents the result.

\newpage
\begin{figure}[ht]
\centering
\input{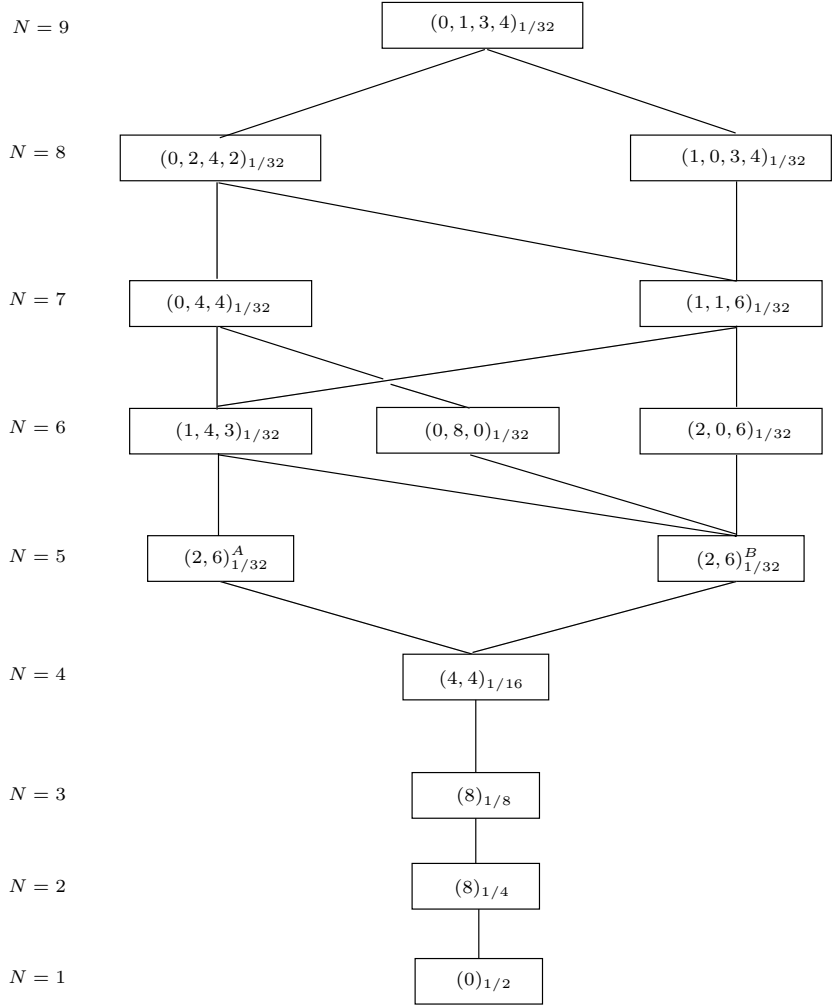}
\caption{{\bf $M$-brane intersections
 with $n=4,5,8$ in 11 dimensions:}{\it  
 Since all
 con\-fi\-gu\-ra\-tions  reduce to $D$-branes in $D=10$ with $n=4,8$
we use $D=10$ labels
 to classify the solutions. For $N=5$ an
extra superscript is added to distinguish between the two sets of labels.
 Subscripts indicate the
 unbroken supersymmetry.}} 
\label{8M-tree}
\end{figure}

Finally we give an example of an $N=9$ configuration in $D=11$.
 Note that $H_5$ and $H_6$ (the lines 5 and 6 in 
 (\ref{N=9,11})) are the $n=8$ pair. $H_5$ may depend on one of the 
coordinates $x_2,x_4,x_5$ or $x_8$, $H_6$ on $x_1,x_3,x_6$ or $x_7$:

{\scriptsize
\begin{equation}  (0,1,3,4)_{1/32}:  \ \ \ \left\{
\begin{array}{c|cc|cc|cc|cc|cc}
\x & \x& \x& - & - & - & - & - & - & - & -  \\
\x & - & - & \x& \x& - & - & - & - & - & -  \\
\x & - & - & - & - & \x& \x& - & - & - & -  \\
\x & - & - & - & - & - & - & \x& \x& - & -  \\
\x & \x& - & \x& - & - & \x& \x& - & \x& -  \\
\x & - & \x& - & \x& \x& - & - & \x& \x& -  \\
\x & \x& - & - & \x& \x& - & \x& - & \x& -  \\
\x & - & \x& \x& - & \x& - & \x& - & \x& -  \\
\x & \x& - & \x& - & \x& - & - & \x& \x& - 
\end{array} \right.
\label{N=9,11}
\end{equation}}

As a final remark, we mention that in the $D$-intersections
 for $n=4,8$ we find  a configuration, $(1,0,0,7)$,
 which  cannot be obtained through dimensional 
 reduction of an intersection of (non-boosted) $M$-branes with $n=4,5,8$.
 As in Section 3, we see that instead 
 the result in $D=11$ has a non-diagonal metric and
 involves a $[5^8]$ configuration  and the Brinkmann wave.
 This can be interpreted as 
 a boosted eight $[5^8]$ intersection in eleven dimensions.
 More explicitly, the configuration $(1,0,0,7)$ ($n=4,8$) in $D=10$ can be written as:
\begin{eqnarray}
ds^2_{10} & = & (H_1H_2H_3H_4H_5H_6H_7H_8H_9)^{-1/2} \{ dt^2 -
(H_1H_2H_4H_5)dx_1^2 \nonumber \\
& &- (H_1H_2H_4H_5H_6H_7H_8H_9)dx_2^2 \nonumber \\
& &- (H_1H_2H_7H_8)dx_3^2 - (H_1H_2H_6H_9)dx_4^2 \nonumber \\
& &- (H_1H_3H_5H_6H_7)dx_5^2 - (H_1H_3H_5H_8H_9)dx_6^2  \nonumber \\
& &- (H_1H_3H_4H_7H_9)dx_7^2 - (H_1H_3H_4H_6H_8)dx_8^2 \\
& &- (H_1H_2H_3H_4H_5H_6H_7H_8H_9)dx_9^2 \} \,,\nonumber \\
e^{-2 \phi} & = & H_1^{-3/2}(H_2H_3H_4H_5H_6H_7H_8H_9)^{1/2}\,, \nonumber \\
A_0 & = & 1- H_1^{-1}\,. \nonumber
\end{eqnarray}
\noindent
Lifted up to eleven dimensions it has the form:

\begin{eqnarray}
ds^2_{11} & = & (H_2H_3H_4H_5H_6H_7H_8H_9)^{-1/3} \{(2-H_1)dt^2 
 - H_1 dx_{10}^2 + 2(1-H_1)dtdx_{10} \nonumber \\
& &- (H_2H_4H_5)dx_1^2 
-(H_2H_4H_5H_6H_7H_8H_9)dx_2^2 - (H_2H_7H_8)dx_3^2 \nonumber \\
& &- (H_2H_6H_9)dx_4^2 - (H_3H_5H_6H_7)dx_5^2 
- (H_3H_5H_8H_9)dx_6^2 \nonumber \\
& &- (H_3H_4H_7H_9)dx_7^2 - (H_3H_4H_6H_8)dx_8^2 \\
& &- (H_2H_3H_4H_5H_6H_7H_8H_9)dx_9^2  \} \,.\nonumber
\end{eqnarray}

It represents an intersection boosted in the direction $x_{10}$
where $H_1$ parametrizes the boost.
If we set $H_1=1$ we recover the  $[5^8]$ $M$-brane intersection. 
If instead we set all $H=1$ except $H_1$, we get the Brinkmann wave in
eleven dimensions.

\end{document}